\documentclass[12pt, aps, pra, twocolumn, superscriptaddress, amsmath, tightenlines, longbibliography,floatfix]{revtex4-2}

\usepackage{dcolumn}
\usepackage{graphicx}
\usepackage{epstopdf}
\usepackage{mathrsfs}
\usepackage{subfigure}
\usepackage{subcaption}
\usepackage{booktabs}
\usepackage{amsmath,amsfonts}
\usepackage{physics}
\usepackage{dsfont}
\usepackage{amstext}
\usepackage{amssymb}
\usepackage{amsbsy}
\usepackage{bbm}
\usepackage{amsthm}
\usepackage{xcolor}
\usepackage{CJK}
\usepackage[colorlinks,urlcolor=blue,linkcolor=blue,citecolor=blue]{hyperref}
\usepackage{soul}

\begin{document}
\title{Quantum Jump Approach for Photosynthetic Energy Transfer with Chemical
Reaction and Fluorescence Loss}

\author{Rui Li}
\affiliation{School of Physics and Astronomy, Applied Optics Beijing Area Major Laboratory, Beijing Normal University, Beijing 100875, China}

\author{Yi Li}
\affiliation{School of Physics and Astronomy, Applied Optics Beijing Area Major Laboratory, Beijing Normal University, Beijing 100875, China}

\author{Kai-Ya Zhang}
\affiliation{School of Physics and Astronomy, Applied Optics Beijing Area Major Laboratory, Beijing Normal University, Beijing 100875, China}

\author{Qing Ai}
\email{aiqing@bnu.edu.cn}
\affiliation{School of Physics and Astronomy, Applied Optics Beijing Area Major Laboratory, Beijing Normal University, Beijing 100875, China}
\affiliation{Key Laboratory of Multistage Spin Physics, Ministry of Education, Beijing Normal University, Beijing 100875, China}

\begin{abstract}
    Recently, the coherent modified Redfield theory (CMRT) has been widely used to simulate the excitation-energy-transfer (EET) processes in photosynthetic systems.
    However, the numerical simulation of the CMRT is computationally expensive when dealing with large-scale systems, e.g. photosystem I (PSI) and II (PSII). On the other hand, the chemical
    reaction and fluorescence loss traditionally treated by the non-Hermitian Hamiltonian approach may result in significantly error in a wide range of parameters.
    To address these issues, we introduce a quantum jump approach (QJA) based on the CMRT to simulate the evolution of photosynthetic complexes including both the chemical
    reaction and fluorescence loss.
    The QJA shows higher accuracy and efficiency in simulating the EET processes.
    The QJA-CMRT approach may provide a powerful tool to design and optimize artificial photosynthetic systems, which benefits future innovation in the field of energy.
\end{abstract}

\maketitle

\section{Introduction}

During the past few decades, substantial progress has been made in elucidating the mechanism of photosynthetic energy transfer.
Nevertheless, simulating the full quantum dynamics in natural photosynthetic complexes (containing 100-300 chromophores) remains computationally  challenging due to the demanding computational resources, e.g., the hierarchical equation of motion~\cite{Tanimura2020JCP}. %{\color{red}\cite{HEOM}}.
Consequently, there exists a critical need to develop computationally-efficient yet theoretically-accurate frameworks capable of modeling the non-Markovian quantum dynamics therein.

In this Letter, we introduce a quantum jump approach (QJA) \cite{Piilo2008PRL,Ai2014NJP} based on the coherent modified Redfield theory (CMRT) \cite{Hwang-Fu2014CP} to simulate the evolution of  photosynthetic complexes under the influence of the chemical reaction (CR) and fluorescence loss (FL).
The CMRT provides a complete description of quantum dynamics of the system's density matrix with the interaction between the system and the environment. 
Based on the CMRT, we use the QJA to transform a set of equations of motion into a stochastic Schr\"{o}dinger equation, and thus significantly reduce the computational resources.
Combining these two approaches, simulations of photosynthetic light harvesting can be achieved with high accuracy and efficiency.

In our QJA-CMRT approach, a non-Hermitian Hamiltonian method, which is equal to the phenomenological density matrix equation demonstrated in Ref.~\cite{Kominis2009PRE}, is used to produce the evolution of the deterministic evolution part of the ensemble. 
As discussions in Ref.~\cite{Kominis2009PRE}, this method is an approximation of the CMRT master equation. 
The deviation between these two theories will be notable when the FL is relatively significant, i.e., {when the ratio of the FL rate to the CR rate is large \cite{Kominis2009PRE,Maeda2008Nature}}. 
However, for the parameter regime of most photosynthetic complexes, these two theories have a good agreement, as shown in Sec.~\ref{sec:simulation}. 
Hence, introducing the non-Hermitian Hamiltonian method into our QJA imposes minimal effects on accuracy while significantly accelerating the simulation for large systems.

This Letter is organized as follows.
In Sec.~\ref{sec:model}, we briefly describe the principle of QJA by dealing with a simplified dimer model.
In Sec.~\ref{sec:simulation}, we apply the QJA to simulate the EET processes in the dimer on account of the CR and FL.
To demonstrate the superiority of the QJA-CMRT approach, we compare the performance with that by the QUTIP \cite{Lambert2023PRR}.
Finally, we summarize the conclusion of this Letter in Sec.~\ref{sec:conclusion}.

\section{Model}
\label{sec:model}

We introduce the QJA-CMRT approach by a dimer model, as schematically illustrated by Fig.~\ref{fig:1(a)}. 
Each site in our model represents a chlorophyll, described as a two-level system. 
Assuming $(\hbar=1)$, the Hamiltonian for this model is given by
\begin{equation}
    H_{S}=\sum_{n=1}^{2}E_{n}|n\rangle\langle n|+J(|1\rangle\langle 2|+|2\rangle\langle 1|),
\end{equation}
where $E_{n}$ denotes the site energy of the $n$th state, 
$J$ represents the electronic-coupling strength between the two states.
Here, $|n\rangle$ is the product state where the $n$th site is on the excited state while the other site is on the ground state, i.e., 
\begin{equation}
    |1\rangle=|e_{1}\rangle\otimes |g_{2}\rangle, \quad |2\rangle=|g_{1}\rangle\otimes|e_{2}\rangle,
\end{equation}
where $|e_{i}\rangle$ ($|g_{i}\rangle$) refers to the excited (ground) state of the $i$th site.
The ground state of this dimer is $|G\rangle=|g_{1}\rangle\otimes|g_{2}\rangle$ with 
the energy $E_{0}=0$.

\subsection{Master Equation Including Chemical Reaction and Fluorescence Loss}

According to the CMRT \cite{Hwang-Fu2014CP}, the evolution of the system is governed by the master equation 
\begin{equation}
    \partial_{t}\rho=-i[H_{S},\rho]-\mathcal{L}_\textrm{pd}(\rho),\label{eq.t1}
\end{equation}
where $\rho$ is the density matrix of the system, and
\begin{equation}
\begin{aligned}
    \mathcal{L}_\textrm{pd}(\rho)=&\sum_{m,n=1}^{2}\frac{R_{mn}}{2}(\{A_{mn}^{\dagger}A_{mn},\rho\}\\
    &-2A_{mn}\rho A_{mn}^{\dagger})
    \label{eq.Lpd}
\end{aligned}
\end{equation}
is the Lindblad term describing the dissipation and dephasing of the system, with $R_{mn}$ being the dissipation rate $(m\neq n)$ and dephasing rate $(m=n)$, $\{A,B\}=AB+BA$ being the anti-commutator, 
and $A_{mn}=|m\rangle\langle n|$ being the jump operator from $|n\rangle$ to $|m\rangle$.
We assume that the CR takes place at site 2.
The CR process could be modeled as the decay of the excitation from $|2\rangle$ to the final state of $|G^{\prime}\rangle$ with the rate $R_\textrm{cr}$.
Moreover, due tothe spontaneous emission of the excited state, the excitation could decay from the single-excitation state $|j\rangle~(j=1,2)$ to the ground state $|G\rangle$ by the FL with the rate $R_\textrm{fl}$.
In Ref.~\cite{Kominis2009PRE}, it has been shown that the traditional treatment of the CR by an exponential decay would result in a significant deviation of theoretical prediction from the experimental observation, and thus it should be treated by the quantum master equation. As a result, we follow the approach to treat the CR and the FL. 
In the master equation, these two effects can be respectively described by the Lindblad term as
\begin{equation}
    \mathcal{L}_\textrm{cr}(\rho)=\frac{R_\textrm{cr}}{2}(\{A_\textrm{cr}^{\dagger}A_\textrm{cr},\rho\}-2A_\textrm{cr}^{\dagger}\rho A_\textrm{cr}),\label{eq.Lcr}
\end{equation}
\begin{equation}
    \mathcal{L}_\textrm{fl}(\rho)=\sum_{n=1}^{2}\frac{R_\textrm{fl}}{2}(\{A_{n}^{\dagger}A_{n},\rho\}-2A_{n}^{\dagger}\rho A_{n}),\label{eq.Lfl}
\end{equation}
where $A_\textrm{cr}=|G^{\prime}\rangle\langle 2|$ and $A_{n}=|G\rangle\langle n|$ are respectively the jump operator of the CR and FL.
Generally speaking, the rates $R_{\text{cr}}$ and $R_{\text{fl}}$ depend on the system parameters, such as the temperature, the spectral density and the level spacing between the two states connected by the quantum jump \cite{Carmichael1993}. 
As the temperature increases, both rates will be enlarged. Moreover, as the level spacing between the two states approaches the maximum of the spectral density, the rate will also be amplified. If the ratio $R_{\text{cr}}/R_{\text{fl}}$ is enlarged, the efficiency will increase. Otherwise, the photosynthesis will be less efficient.
Taking the CR and FL into consideration, the total master equation of the system can be written as
\begin{equation}
    \partial_{t}\rho=-i[H_{S},\rho]-\mathcal{L}_\textrm{pd}(\rho)-\mathcal{L}_\textrm{cr}(\rho)-\mathcal{L}_\textrm{fl}(\rho).\label{eq.7}
\end{equation}

\subsection{Quantum Jump Approach}

In this subsection, we provide an overview of the QJA \cite{Piilo2008PRL,Ai2014NJP} to unravel the master equation.
In the Markovian regime, all decay rates remain positive, whereas in the non-Markovian regime, some decay rates may become negative during some time intervals.
Consequently, the decay channels are classified into two categories, i.e., positive and negative, denoted by $j_+$ and $j_-$, respectively. 
Their corresponding decay rates (jump operators) are $R_{j_+}(t)>0~(A_{j_+}(t))$ and $R_{j_-}(t)<0~(A_{j_-}(t))\label{eq.Rj_-}$.
Moreover, as the rates of CR and FL consistently remain positive, they are thereby assigned to the positive channels $j_+$ at all times.

In order to calculate the system's evolution, we consider $M$ independent samples evolving in parallel, and the system's actual dynamics can be obtained by averaging over these $M$ trajectories.
The density matrix of the system could be written as
\begin{equation}
    \rho(t)=\frac{1}{M}\sum_{\alpha=1}^{M}|\psi_{\alpha}(t)\rangle\langle\psi_{\alpha}(t)|,\label{eq.rho-t}
\end{equation}
where $|\psi_{\alpha}(t)\rangle$ is the state of the $\alpha$th sample at time $t$.
The initial state of the system is assumed to be a maximum-mixed state, i.e., $\rho(0)=(|1\rangle\langle 1|+|2\rangle\langle 2|)/2$.
Up to a normalization factor, after a sufficiently-small time step, the state $|\psi_{\alpha}(t)\rangle$ will  deterministically evolve into
\begin{align}
    |\phi_{\alpha}(t+\delta t)\rangle&=\exp(-iH\delta t)|\psi_{\alpha}(t)\rangle\nonumber\\
    &\approx(1-iH\delta t)|\psi_{\alpha}(t)\rangle,\label{eq.9}
\end{align}
with the non-Hermitian Hamiltonian $H$ 
\begin{align}
    H=&H_{S}-i\sum_{j_+}\frac{R_{j_+}(t)}{2}A_{j_+}^{\dagger}(t)A_{j_+}(t)\nonumber\\
    &-i\sum_{j_-}\frac{R_{j_-}(t)}{2}A_{j_-}^{\dagger}(t)A_{j_-}(t).\label{eq.10}
\end{align}
After normalization, the state is $|\psi_{\alpha}(t+\delta t)\rangle=|\phi_{\alpha}(t+\delta t)\rangle/\||\phi_{\alpha}(t+\delta t)\rangle\|$.
As a result of the decay channels, the deterministic evolution has a certain probability to be interrupted by quantum jumps to the other states.
For the positive channels, the system can jump to the state $A_{j_+}(t)|\psi_{\alpha}(t)\rangle/\|A_{j_+}(t)|\psi_{\alpha}(t)\rangle\|$, with the probability
\begin{align}
    P_{\alpha}^{j_+}(t)=&\langle\psi_{\alpha}(t)|A_{j_+}^{\dagger}(t)A_{j_+}(t)|\psi_{\alpha}(t)\rangle \nonumber\\
    &\times R_{j_+}(t)\delta t\label{eq.11}
\end{align}
during a sufficiently-short time interval $\delta t$.
For the negative channels, the system can jump to the state
\begin{equation}
    |\psi_{\alpha}(t)\rangle=\frac{A_{j_-}(t)|\psi_{\alpha^{\prime}}(t)}{\|A_{j_-}(t)|\psi_{\alpha^{\prime}}(t)\rangle\|},
    \label{12}
\end{equation}
where the jump operator is
\begin{equation}
    A_{j_-}(t)=|\psi_{\alpha^{\prime}}(t)\rangle\langle\psi_{\alpha}(t)|.
    \label{13}
\end{equation}
The corresponding probability is
\begin{align}
    P_{\alpha\alpha^{\prime}}^{j_-}(t)=&\langle\psi_{\alpha^{\prime}}(t)|A_{j_-}^{\dagger}(t)A_{j_-}(t)|\psi_{\alpha^{\prime}}(t)\rangle \nonumber\\
    &\times \frac{M_{\alpha^{\prime}}(t)}{M_{\alpha}(t)}|R_{j_-}(t)|\delta t.\label{eq.14}
\end{align}

The decay processes involving positive and negative channels can be interpreted as follows.
Consider a given channel $j$. During some time interval, when $R_{j}>0$, the quantum jump to proceed in the postive direction, i.e., $|\psi_{\alpha}\rangle\rightarrow|\psi_{\alpha^{\prime}}\rangle=A_{j}|\psi_{\alpha}\rangle/\|A_{j}|\psi_{\alpha}\rangle\|$.
However, during the other time interval, when $R_{j}<0$, the quantum jump occurs in the reverse direction, i.e., $|\psi_{\alpha}\rangle\leftarrow|\psi_{\alpha^{\prime}}\rangle$. In other words, the positive jump losses the coherence, while the negative jump recovers the coherence and thus the memory. 

Using Eqs.~(\ref{eq.10})-(\ref{eq.14}), the density matrix of the system at $t+\delta t$ reads
\begin{widetext}
\begin{eqnarray}
    \bar{\rho}(t+\delta t)&=&\sum_{\alpha}\frac{M_{\alpha}}{M}\Bigg[\Bigg(1-\sum_{j_+}P_{\alpha}^{j_+}(t)-\sum_{j_-}P_{\alpha\alpha^{\prime}}^{j_-}(t)\Bigg)\frac{|\phi_{\alpha}(t+\delta t)\rangle\langle\phi_{\alpha}(t+\delta t)|}{\||\phi_{\alpha}(t+\delta t)\rangle\|^{2}}\nonumber\\
    &&+\sum_{j_+}P_{\alpha}^{j_+}(t)\frac{A_{j_+}(t)|\psi_{\alpha}(t)\rangle\langle\psi_{\alpha}(t)|A_{j_+}^{\dagger}}{\langle\psi_{\alpha}(t)|A_{j_+}^{\dagger}A_{j_+}(t)|\psi_{\alpha}(t)\rangle}+\sum_{\alpha^{\prime},j_-}P_{\alpha\alpha^{\prime}}^{j_-}(t)|\psi_{\alpha^{\prime}}\rangle\langle\psi_{\alpha^{\prime}}|\Bigg].\label{eq.15}
\end{eqnarray}
\end{widetext}
When the ensemble is sufficiently large, i.e., $M\gg1$, Eq.~(\ref{eq.15}) fulfills Eq.~(\ref{eq.7}), {as proven in the Supplementary Material.}

In the case of {$N$} sites, the QJA obtains the state evolution in $\delta t$ through computing the equations about state vector $|\psi_{\alpha}(t)\rangle$,
whose complexity is linearly related to $N$.
Comparing to the direct unravelling the master Eq.~(\ref{eq.7}) by the QUTIP, whose complexity is related to $N^{2}$, the QJA may show some superiority in obtaining the same result.
The explicit dependency of the running time $t$ on $N$ for these two methods are compared in Sec.~\ref{sec:simulation}. 
Therefore, the QJA is a powerful tool to simulate the evolution of our photosynthetic systems.

\section{Numerical Simulation and Discussion}
\label{sec:simulation}

In this section, we utilize the QJA-CMRT to simulate the EET processes in several different model systems, including the dimer model and a stochastic system based on PSI. 
We have simulated the evolution of the systems using CMRT master equation supported by QUTIP and CMRT-QJA, labeled by QUTIP and QJ below, respectively.

\subsection{The Dimer Model}

\begin{figure*}
    \centering
    \includegraphics[width=0.9\textwidth]{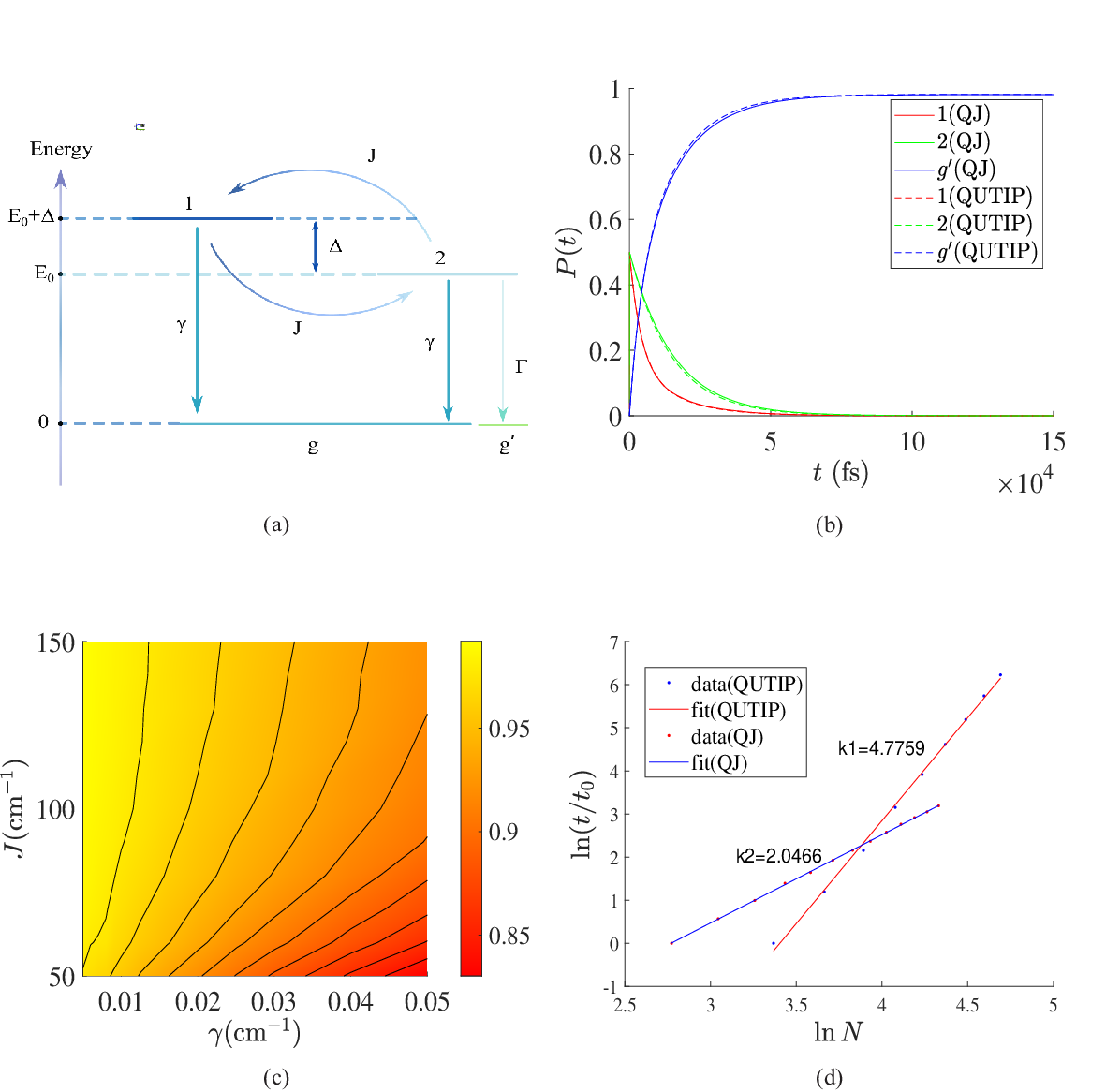}
    \caption{(a) An energy diagram of the dimer system.
The two single-excitation states are $\vert1\rangle$ and $\vert2\rangle$. The ground state and the product state of the CR are $\vert g\rangle$ and $\vert g^\prime\rangle$, respectively.\label{fig:1(a)}
(b) The population dynamics of the dimer system are respectively simulated by the QJA (solid line) and the QUTIP (dashed line).\label{fig:1(b)}
(c) The dependence of efficiency $\eta$ on the electronic coupling $J$ and the FL rate $\gamma$.
The energy gap $\Delta$ and the CR rate are fixed at $763$~cm$^{-1}$ and $1.33$~cm$^{-1}$.\label{fig:1(c)}
(d) The scaling of the computational time $t$ against the number of sites $N$ by the two approaches.\label{fig:1(d)}}
\end{figure*}

The dimer model is schematically illustrated in Fig.~\ref{fig:1(a)}. The Hamiltonian in the site basis reads
\begin{equation}
    H=\left[\begin{matrix}
        E_{0}&J\\
        J&E_{0}+\Delta
    \end{matrix}\right].
\end{equation}
In practical, the system is initially in the maximum-mixed state of all sites, due to the wide spread of the spectrum of sunlight. 
In our simulation, we adopt the same parameters as the EC-A1 and EC-B1 sites of PSI \cite{Damjanovic2002JPCB}, i.e., 
$E_0=13201$~cm$^{-1}$ and $\Delta=763$~cm$^{-1}$. 
The rates of CR and FL are respectively $\Gamma\sim1.33$~cm$^{-1}$ and $\gamma\sim8.33\times10^{-3}$~cm$^{-1}$~\cite{Kruger2017JPB}. 
Fig.~\ref{fig:1(b)} illustrate the population dynamics of the dimer model simulated by the QJA and QUTIP, which are represented by solid and dashed lines respectively. 
Clearly, the results by the two methods are in perfect agreement, which strongly support the accuracy of the QJA. 
In addition, it is worth noting that the equilibrium time of the system varies in $\sim 10$-$25$ps, being close to the EET time scale in Ref.~\cite{Kruger2017JPB}, following the variation of electronic coupling $J$, but have merely no response to the change of $\gamma$ in a wide parameter regime from $10^{-3}\sim10^{-1}~\text{cm}^{-1}$. 
This fact indicates that couplings between chromophores play a more crucial role in the evolution than the correlations between the system and the environment.

The efficiency $\eta$ against $J$ and $\gamma$ is investigated by using QJA as shown in Fig.~\ref{fig:1(c)}. 
We could find that the efficiency $\eta$ rises along with increasing $J$, while it monotonically falls as $\gamma$ increases, which is in agreement with our physical intuition, that the electronic coupling $J$ (the system-bath coupling $\gamma$) promotes (suppresses) the EET.

\subsection{The $N$-dependency of the computational time of QJA and QUTIP}

In this subsection, we compare the dependency of the computational time $t$ on the number of the sites $N$ of the QJA and QUTIP. 
We assume that the relation between $t$ and $N$ can be characterized by a power-law expression as
\begin{equation}
    t=aN^{b},
\end{equation}
where $a$ and $b$ are constants. 
Generally speaking, the parameter $a$ is related to the performance of the computer and the size of the ensemble, while $b$ is an intrinsic property of the algorithm.
Thus, we focus on the value of $b$.
In Fig.~\ref{fig:1(d)}, we compare the performance of the QJA and QUTIP. 
In order to better show the scaling of $N$, we have normalized the simulation data by using the first computational time of each method as 1.
The multiple-site model using in our simulation is based on the PSI, in which $N$ sites are randomly selected while the couplings between any two sites therein are retained as the original PSI.
According to our numerical simulations which are not shown here, for a given $N$ the conclusion about the computation times by the two approaches will not be changed if we choose different sets of sites for simulation.
The size $N$ in our simulation varies from 10 to 100, with a step of 5. As illustrated in Fig.~\ref{fig:1(d)}, the trend can be well fitted with a linear model, with $R$-squared as $R^{2}_{\text{QJA}}=0.9998$ and $R^{2}_{\text{QUTIP}}=0.9981$.
The slope $b$ of QJA is $2.0$ is much smaller than that of QUTIP, i.e., $4.8$. 
This observation implies that the QJA is much more efficient than the QUTIP in simulating photosynthetic systems, especially when applying to the large-scale systems like PSI and PSII.

\section{Conclusion}
\label{sec:conclusion}

In summary, we have introduced the QJA based on the CMRT to simulate the EET processes in photosynthetic systems with the CR and FL.
The QJA-CMRT approach is capable of accurately simulating the EET processes in large-scale systems, such as PSI and PSII, while maintaining a high level of computational efficiency.
Compared with the CMRT unravelled by the QUTIP, the QJA-CMRT approach has a much lower computational complexity, which has an $N^{2.0}$ dependency on the number $N$ of sites.
In previous studies, the investigation has been focused on small-scale photosynthetic systems for its simplicity to simulate.
With the QJA-CMRT approach, it is possible to investigate the spacial and energetic structure of large-scale systems, such as PSI and PSII, which is a significant step towards the design and optimization of artificial photosynthetic systems.

We acknowledge valuable discussions with Jun Wang. 
This work is supported by Innovation Program for Quantum Science and Technology under Grant No.~2023ZD0300200, the National Natural Science Foundation of China under Grant No.~62461160263, Beijing Natural Science Foundation under Grant No.~1202017, and Beijing Normal University under Grant No.~2022129.

\bibliography{total}

@PREAMBLE{
 "\providecommand{\noopsort}[1]{}"
 # "\providecommand{\singleletter}[1]{#1}%"
}

@article{Kruger2017JPB,
  title={The role of energy losses in photosynthetic light harvesting},
  author={Kr{\"u}ger, T.-P.-J. and van Grondelle, R.},
  journal={J. Phys. B},
  volume={50},
  number={13},
  pages={132001},
  year={2017},
  publisher={IOP Publishing},
  doi = {10.1088/1361-6455/aa7583},
  url = {https://dx.doi.org/10.1088/1361-6455/aa7583},
}

@article{Damjanovic2002JPCB,
  title={Chlorophyll excitations in photosystem {I} of Synechococcus elongatus},
  author={Damjanovi{\'c}, A. and Vaswani, H.-M. and Fromme, P. and Fleming, G.-R.},
  journal={J. Phys. Chem. B},
  volume={106},
  number={39},
  pages={10251--10262},
  year={2002},
  publisher={ACS Publications},
  doi = {10.1021/jp020963f},
  URL = {https://doi.org/10.1021/jp020963f},
}

@article{Lambert2023PRR,
  title = {QuTiP-BoFiN: A bosonic and fermionic numerical hierarchical-equations-of-motion library with applications in light-harvesting, quantum control, and single-molecule electronics},
  author = {Lambert, N. and Raheja, T. and Cross, S. and Menczel, P. and Ahmed, S. and Pitchford, A. and Burgarth, D. and Nori, F.},
  journal = {Phys. Rev. Res.},
  volume = {5},
  issue = {1},
  pages = {013181},
  numpages = {18},
  year = {2023},
  month = {Mar},
  publisher = {American Physical Society},
  doi = {10.1103/PhysRevResearch.5.013181},
  url = {https://link.aps.org/doi/10.1103/PhysRevResearch.5.013181}
}

@article{Hwang-Fu2014CP,
title = {{A coherent modified Redfield theory for excitation energy transfer in molecular aggregates}},
journal = {Chem. Phys.},
volume = {447},
pages = {46-53},
year = {2015},
issn = {0301-0104},
doi = {https://doi.org/10.1016/j.chemphys.2014.11.026},
url = {https://www.sciencedirect.com/science/article/pii/S0301010414003358},
author = {Yu-Hsien Hwang-Fu and Wei Chen and Yuan-Chung Cheng},
}

@article{Tanimura2020JCP,
    author = {Tanimura, Y.},
    title = "{Numerically  ``exact" approach to open quantum dynamics: {The} hierarchical equations of motion (HEOM)}",
    journal = {J. Chem. Phys.},
    volume = {153},
    number = {2},
    pages = {020901},
    year = {2020},
    month = {07},
    issn = {0021-9606},
    doi = {10.1063/5.0011599},
    url = {https://doi.org/10.1063/5.0011599}
}

@article{Maeda2008Nature,
	year = 2008,
	volume = {453},
	number = {7193},
	pages = {387},
                 numpages = {4},
	author = {Maeda, Kiminori and Henbest, Kevin B. and Cintolesi, Filippo and Kuprov, Ilya and Rodgers, Christopher T. and Liddell, Paul A. and Gust, Devens and Timmel, Christiane R. and Hore, P. J.},
	title = {Chemical compass model of avian magnetoreception},
	journal = {Nature},
       doi={10.1038/nature06834}
}

@article{Kominis2009PRE,
    author = {Kominis, I. K.},
    issn = {15393755},
    journal = {Phys. Rev. E},
    month = {nov},
    number = {5},
    pages = {56115},
    publisher = {American Physical Society},
    title = {{Quantum Zeno effect explains magnetic-sensitive radical-ion-pair reactions}},
    volume = {80},
    year = {2009},
    url={http://dx.doi.org/10.1103/PhysRevE.80.056115},
}

@article{Piilo2008PRL,
  title = {{Non-Markovian} Quantum Jumps},
  author = {Piilo, Jyrki and Maniscalco, Sabrina and H\"ark\"onen, Kari and Suominen, Kalle-Antti},
  journal = {Phys. Rev. Lett.},
  volume = {100},
  issue = {18},
  pages = {180402},
  numpages = {4},
  year = {2008},
  month = {May},
  publisher = {American Physical Society},
  doi = {10.1103/PhysRevLett.100.180402},
  url = {https://link.aps.org/doi/10.1103/PhysRevLett.100.180402}
}

@ARTICLE{Ai2014NJP,
   author       = " Q. Ai and Y. J. Fan and B. Y. Jin and Y. C. Cheng",
   title        = "An efficient quantum jump method for coherent energy transfer dynamics in photosynthetic systems under the influence of laser fields",
   journal      = "New J. Phys.",
   volume       = "16",
   pages        = "053033",
   year         = "2014",
url={http://iopscience.iop.org/1367-2630/16/5/053033},
}

@BOOK{Carmichael1993,
   author       = "H. J. Carmichael",
   year         = "1993",
   title        = "An Open Systems Approach to Quantum Optics",
   publisher    = "Spring-Verlag",
   address        = "Gemany",
}
\end{document}